\documentclass{l4dc2024}

\usepackage{wrapfig}
\usepackage{times}
\usepackage{lipsum}
\usepackage{amsfonts} 
\usepackage{mathtools} 
\usepackage{subcaption}
\usepackage{graphicx}
\usepackage{caption}

\newcommand{\B}[1]{\if#1\relax\bm
{#1}\else\mathbf{#1}\fi} 
\newcommand{\C}[1]{\mathcal{#1}}
\newcommand{\BB}[1]{\mathbb{#1}}

\title[\textit{In vivo} learning-based control]{\textit{In vivo} learning-based control of microbial populations density\\ in bioreactors}
\usepackage{times}

\author{%
 \Name{Sara Maria Brancato} \Email{saramaria.brancato@unina.it}\\
 \addr University of Naples Federico II
 \AND
 \Name{Davide Salzano} \Email{davide.salzano@unina.it}\\
 \addr Scuola Superiore Meridionale%
 \AND
 \Name{Francesco {De Lellis}} \Email{francesco.delellis@unina.it}\\
 \addr University of Naples Federico II%
 \AND
 \Name{Davide Fiore} \Email{davide.fiore@unina.it}\\
 \addr University of Naples Federico II%
 \AND
 \Name{Giovanni Russo*} \Email{giovarusso@unisa.it}\\
 \addr University of Salerno%
 \AND
 \Name{Mario {di Bernardo}*} \Email{mario.dibernardo@unina.it}\\
 \addr University of Naples Federico II%
}

\begin{document}

\maketitle

*Corresponding authors

\begin{abstract}%
A key problem toward the use of microorganisms as bio-factories is reaching and maintaining cellular communities at a desired density and composition so that they can efficiently convert their biomass into useful compounds. Promising technological platforms for the \textit{real time}, scalable control of cellular density are bioreactors. In this work, we developed a learning-based strategy to expand the toolbox of available control algorithms capable of regulating the density of a \textit{single} bacterial population in bioreactors. Specifically, we used a \textit{sim-to-real} paradigm, where a simple mathematical model, calibrated using a few data, was adopted to generate synthetic data for the training of the controller. 
The resulting policy was then exhaustively tested \textit{in vivo} using a low-cost bioreactor known as Chi.Bio, assessing performance and robustness.
In addition, we compared the performance with more traditional controllers (namely, a PI and an MPC), confirming that the learning-based controller exhibits similar performance \textit{in vivo}. Our work showcases the viability of learning-based strategies for the control of cellular density in bioreactors, making a step forward toward their use for the control of the composition of microbial consortia.
\end{abstract}


\section{Introduction}

\begin{figure}
    \centering
    \includegraphics[width=0.4\linewidth]{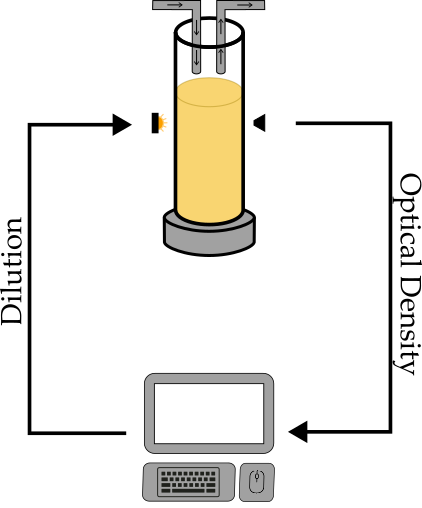}
    \caption{An automated cell growth setup: the cell density at a given time is estimated via optical density measures, while a computer automatically implements a control law able to regulate the dilution  (hence the density inside the chamber) by adding fresh media and discarding the waste.}
    \label{fig:Chi.Bio}
\end{figure}

Microorganisms, such as bacteria and yeast, have been used in industry as efficient, low-waste bio-factories able to convert nutrients into useful proteins or chemicals \citep{brenner2008engineering,satyanarayana2009yeast,su2020bacillus,jullesson2015impact,choi2018recombinant,hug2020bacteria}. 
This is made possible by engineering \textit{de-novo} synthetic circuits into cells or combining the natural bio-processing capabilities of different organisms.
In this context, an important issue is how to employ cell resources to efficiently transform biomass into protein production while preventing the accumulation of toxic by-products \citep{mauri2020enhanced,tian2020titrating,xu2018production,lv2019coupling}.
Using bioreactors, it is possible to reach and maintain a desired cell density in the growth environment, to create the optimal growth conditions for the bio-production of a given chemical.
Figure \ref{fig:Chi.Bio} presents an example of automated control architecture applied to cell growth regulation. Specifically, by modulating the dilution with the introduction of new nutrients in the growth environment, it is possible to adjust in \textit{real time} the density of the culture.  
To do so, an external controller can be designed to run on a computer and automatically regulate the density of the cells in bioreactors, by evaluating the error between the measure of the density inside the chamber and the desired density level to be achieved.
Various strategies exist for regulating cell populations within a chamber, including those that manipulate dilution rates in chemostats \citep{de2003feedback}, while others leverage genetic interventions of the cell strains, and utilize different control inputs such as lights \citep{gutierrez2022dynamic} or various nutrients \citep{treloar2020deep}.
From a control design perspective, existing approaches utilize traditional controllers like PIs \citep{kusuda2021reactor}, non-linear piecewise smooth methods, or gain scheduling state feedback strategies \citep{fiore2021feedback}. Some also harness computational capabilities to derive control laws incorporating constraints, either through mechanistic models \citep{bertaux2022enhancing,aditya2021light,zhu2000model} or entirely through data-driven methods utilizing reinforcement learning \citep{treloar2020deep}.

Recent developments in quantitative systems and synthetic biology have led to the increased adoption of compact and cost-effective bioreactors, such as those explored by \cite{bertaux2022enhancing,steel2019chi,wong2018precise}. These bioreactors offer integrated control equipment and multiple sensors in a unified platform, enabling precise manipulation of environmental conditions for extended periods of time in microbial cultures, making them highly attractive for controlling microbial consortia.
Among the different low-cost, open-source bioreactor platforms available for the rapid prototyping of novel microbial communities for bio-production, the Chi.Bio \citep{steel2019chi} offers the possibility to have a controlled, static environment in which culture parameters such as nutrient availability and temperature can be regulated, and includes the ability to frequently measure cellular density and bulk fluorescence, and the possibility of optogenetic actuation.
This platform makes use of a PI controller for the \textit{real time} control of the density of the cells in the culture vial. Although this controller is able to stabilize and maintain the density of a cellular population to a desired fixed value, the optimal tuning of the controller gains requires accurate knowledge of the controlled system.

An alternative approach to overcome the need for an accurate, well-calibrated mathematical model is to leverage learning-based control methods to learn the policy by directly interacting with the system. As proposed by \citep{treloar2020deep}, a suitable control law can be learned within 24 hours using five parallel bioreactors. However, this approach was only validated \textit{in-silico} and doesn't consider the stochasticity and variability of the biological process, displaying the necessity for additional investigations into the feasibility of such approaches. Indeed, the learning process can be ``sample inefficient'', requiring long times and a large number of experimental data to learn the policy, which could hinder its use in biology (see \cite{BUSONIU20188,MC1}).
A possible solution to learn a control policy without the need for a large set of experimental data comes from the \emph{sim-to-real} approach, where the control policy is learnt on simulated environments and subsequently exported to the real system \citep{pmlr-v78-rusu17a, tan2018sim, james2017transferring}. This can be particularly challenging in applications to biological systems as they evolve and grow, and are characterized by cell-to-cell variability, uncertainties and other disturbances that are cumbersome to accurately capture in synthetic mathematical models. Hence, a key open problem is to understand if and how learning-based controllers trained using a sim-to-real approach can be effectively deployed \textit{in vivo} to control bacterial populations.

In this work, we address this problem by developing a learning-based controller for the regulation of cellular density to a desired value in a bioreactor. In line with the \emph{sim-to-real} approach, the control law is learnt by interacting with synthetically generated data. These data are generated from a simple model capturing the main features of the growth dynamics. 
Note that, even though partial knowledge of the system's dynamics is needed, 
a coarse calibration of the parameters, obtained using a few open-loop experiments, is enough to generate the data required for the training of the control algorithm.
We show via a set of exhaustive \emph{in vivo} experiments that the \textit{sim-to-real} gap can be filled and that the control performance learnt using a simple, inaccurate model can be transferred to real experiments carried out using a bioreactor. 
We benchmark our controller in terms of performance and robustness against the on-board PI controller in the Chi.Bio and a Model Predictive Controller that was developed for the sake of comparison taking advantage of the simple model used to generate synthetic data for the learning-based controller.

\section{Control problem formulation}
\label{sez:sec_control_p}
We consider the growth dynamics of a bacterial species inside a microbial culture chamber described as a continuous time dynamical system of the form:
\begin{equation}\label{eq:continuous_dynamical_system}
\begin{split}
    \dot{x}_{t} & = f(x_{t}, u_{t}),
    \quad x_{0} = \tilde{x}_{0}, \\
    y_t & = \alpha \, x_t,
\end{split}
\end{equation}
where $x_{t} \in \C{X}$ is the concentration of bacteria in the microbial culture chamber at time $t$, with
$\C{X} \subseteq \mathbb{R}$ being the state space, 
$\tilde{x}_0 \in \C{X}$ is the initial concentration, 
$u_t \in \C{U} $ is the control input or \emph{pump rate} delivered as an exogenous injection of fresh growth media in the microbial chamber, with $\C{U} \subseteq \mathbb{R}$ being the set of feasible inputs, $f : \C{X} \times \C{U} \rightarrow \C{X}$ is the vector field defining the system dynamics, and the output $y \in [0,1]$ is the optical density (OD) measured by the platform, expressed in arbitrary units.
For the sake of simplicity, we assume that $\alpha$ is equal to 1, and therefore from now on we will equivalently refer to either bacterial concentration $x$ or optical density $y$.

To take into account the technological constraints of common microbiology platforms, we consider the case where the control input can only be applied at fixed discrete time steps.
Therefore, we design our control strategy starting from the following discrete time dynamical system:
\begin{equation}\label{eq:dynamical_system}
    x_{t_{k+1}} = x_{t_k} + \int_{t_k}^{t_{k+1}} f(x_\tau, u_{t_k}) \,d\tau,
\quad x_{t_0} = \tilde{x}_{0},
\end{equation}
where $t_{k} \in \BB{N}_{\ge 0}$ is the discrete time, and $u_{t_k}$ is a piecewise constant function defining the constant \emph{pump rate} applied in the time interval $[t_k, t_{k+1})$ to the system dynamics \eqref{eq:continuous_dynamical_system}.
Moreover, notice that when $u_{t_k}\neq 0$, that is when fresh media is pumped into the chamber, the experimental platform automatically expels some of the fluid from the chamber with a rate greater then the input rate, in order to avoid overflow of the media from the vessel. 
Considering the following assumptions:
\begin{itemize}
    \item A1. The concentration x is quantified through OD measures 
    \item A2. The measures are collected every minute
    \item A3. The control input, i.e. the pump rate is limited to avoid overflow
\end{itemize}
 the \emph{control goal} is to regulate at steady state the bacterial concentration $x$ in the chamber to some desired value $\bar{x} \in [0.2, 1]$, which corresponds to the operating condition for which the cells are at exponential growth, where protein production is facilitated.

%

\subsection{Stating the learning-based control problem}
Following \citep{pmlr-v168-lellis22a, pmlr-v211-de-lellis23a}, the previous control goal can be reformulated as a learning-based control problem. 
Specifically, we want to learn the control policy $\pi : \C{X} \rightarrow \C{U}$ 
to solve the following optimal control problem over a finite time horizon $t_N \in \BB{N}_{>0}$: 
\begin{subequations}\label{eq:rl_problem_statement}
\begin{align}
    \max_{\pi}& \ \ \BB{E}[J^{\pi}],\\ 
    \text{s.t.}
    & \ \ x_{t_{k+1}} = x_{t_k} + \int_{t_k}^{t_{k+1}} f(x_\tau, u_{t_k}) \,d\tau,
        \quad t_k \in \{ 0, \dots, t_{N-1} \},\\
    &\ \ u_{t_k} = \pi(x_{t_k}),  
        \quad t_k \in \{ 0, \dots, t_{N-1} \},\\
    &\ \ x_{t_0} \ \text{given},
\end{align}
\end{subequations}
where the objective function is a \emph{discounted cumulative reward} defined as:
\begin{equation} \label{eq:cumulative_reward}
J^{\pi} = r_{t_N}(x_{t_N}) + \sum_{k=0}^{N-1} \gamma^{k} r(x_{t_k}) ,
\end{equation}
with $r : \C{X}  \rightarrow \BB{R}$ being the \emph{reward} received by the learning agent, $\gamma$ a forgetting factor equal to $0.99$, and $r_{t_N}:\C{X} \rightarrow \BB{R}$ being the \emph{final reward}.
In particular, the reward function is formulated as a distance-like function between the bacterial density in the chamber and a given reference setpoint $\bar{x}$ as follows:
\begin{equation}\label{eq:reward_function}
    r(x_{t_k}) = - (x_{t_k} - \bar{x})^2 ,
\end{equation}
which steers the learning agent towards achieving and maintaining the bacterial density to the reference setpoint value $\bar{x}$ . 

\section{Control design and validation}

To solve the above learning problem and thus regulate the density of the bacterial population in a bioreactor, we designed a Deep Q-Learning algorithm leveraging the so-called \textit{sim-to-real} approach.
Specifically, as a test-bed species, we utilized the \textit{Escherichia coli} (\textit{E. coli}) strain designed by \cite{gardner2000construction} embedding a plasmid that implements a genetic toggle-switch (i.e. a reversible bistable memory mechanism).

In this section, we illustrate the three-step pipeline we used to develop our control algorithm (Figure \ref{fig:Pipeline_Deploy}).
%
Specifically, first, we chose and calibrated a dynamical model able to capture the growth dynamics of the microorganisms. Then, the mathematical model was used to generate synthetic data for the training of the neural network. Finally, the trained network was deployed \textit{in vivo} to control the density of the population inside the bioreactor.

\subsection{Modeling the microbial growth simulator}

The production of synthetic data for the training of the model requires the choice and parametrization of a mathematical model capturing the main dynamical features of the growth of the bacteria. An established model for the description of exponential growth of bacteria in bioreactors \citep{monod1949growth} can be written as:
\begin{equation} \label{eq:simple_dynamics}
    \dot{x}_t = \left( \mu - \frac{u_t}{\tau} \right) x_t ,
\end{equation}
where $x$ is the density of the cellular population, $\mu$ is the growth rate of the population, $\tau$ is a scaling factor,
and $u$ represents our control input (i.e. the dilution rate applied by modulating the speed of the pump carrying fresh media into the reactor). 
All the quantities in the above model are adimensional. Indeed, the measured optical density takes values between 0 and 1, which correspond to the absence and abundance of bacteria in the chamber, respectively, and are calibrated at the beginning of the experiments. 
To parametrize this model we carried out a single open-loop experiment growing the bacteria in the Chi.Bio at different values of the dilution rate changed randomly every 30 minutes. 
All the experiments, unless otherwise stated, were performed at 37 $^o$C in luria broth media supplemented with $50\, \mu g/\mu L$ Kanamycin and $1\, mM$ Isopropyl $\beta$-D-1-thiogalactopyranoside (IPTG). 

The values of $\mu$ and $\tau$ were estimated from experimental data using a least square estimator in MATLAB and validated via open-loop experiments. 
In these experiments, cells were grown for 60 minutes. Subsequently, the cell culture was diluted using the maximum available dilution rate of $0.02 mL/s$ until the optical density fell below 0.3. Finally, the dilution rate was randomly changed every 30 minutes. 
Figure \ref{fig:Pipeline_Deploy} (bottom left panel) shows a comparison between the data generated by the model (in blue) and the real data recorded from the Chi.Bio (in red). Note that the model can effectively capture both the dynamics of the exponential growth of the population and the effects of dilution. 
However, as expected the prediction of the system trajectories is not very accurate, as demonstrated by the relatively high mean squared error between the estimated data and the recorded data (with a percentage mean square error (PMSE) equal to $6\%$). The question is now whether using such a simple model can be effective when generating synthetic data for the design of a learning-based controller to be used \textit{in vivo}.

\subsection{Training and deployment of the learning-based controller}
%

\begin{figure}
    \centering
    \includegraphics[width=1\linewidth]{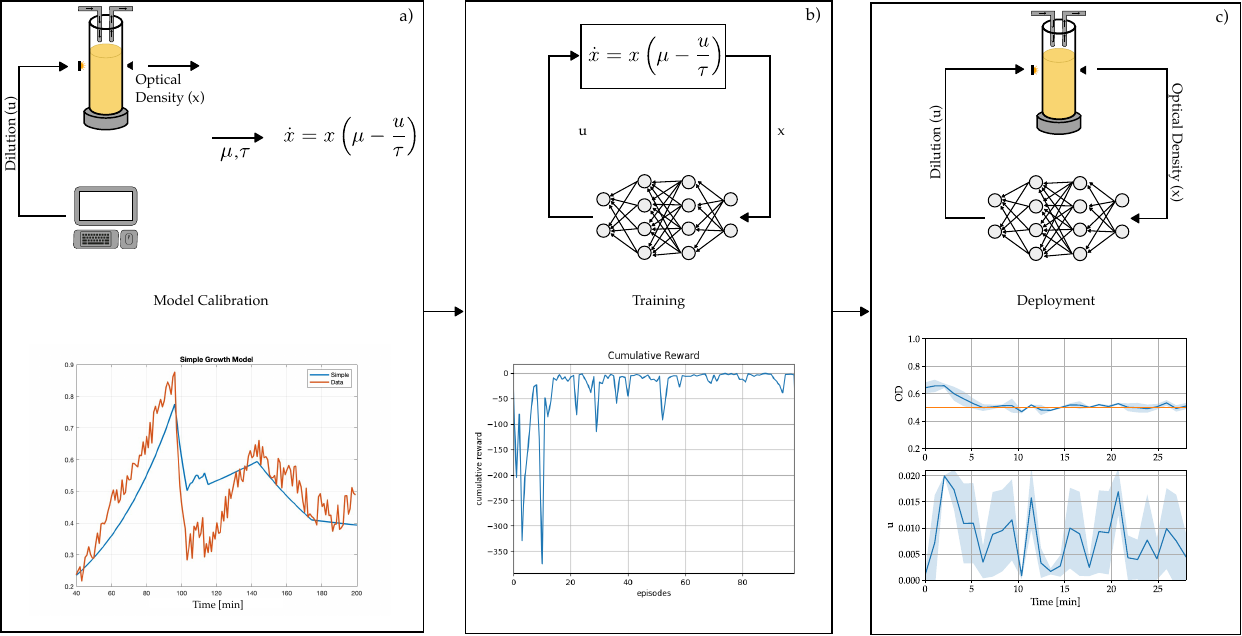}
    \caption{Sim-to-real pipeline. Panel a) shows the selection and calibration of a dynamic model describing the microorganisms' growth; the bottom panel depicts the comparison between the fitted model and the data collected with the Chi.Bio. 
    Panel b) displays the training phase: the mathematical model was employed to produce synthetic data for the training of the neural network. The bottom panel reports the cumulative reward progress in 100 episodes. 
    Panel c) represents how the trained network is employed to regulate the cell population density within the bioreactor. Bottom panel shows the time evolution of the OD controlled by the DQN-based control algorithm and the corresponding control input sent to the pump. Solid lines represent the average evolution of state and input, while the shaded areas represent the standard deviation obtained over three \textit{in vivo} experiments. The orange dashed line indicates the set point $\bar{x}$ of 0.5. }
    \label{fig:Pipeline_Deploy}
\end{figure}

We implement a DQN algorithm \citep{mnih2015human} in which a neural network approximates the optimal action-value function (see \cite{watkins1992q}). Specifically, the neural network is used to estimate the action $u$ based on the OD measure $x$ and the desired reference values $\bar{x}$ of the OD, which are the neural network inputs. The training is performed by using synthetic data generated by the simplified mathematical model \eqref{eq:simple_dynamics}, whilst $\bar{x}$ is randomly drawn with uniform distribution from the discrete set $\{0.2,\, 0.3,\, \dots\, ,\, 0.9,\, 1\}$ at each episode.
The possible control actions, $u$, are 17 discrete values taken uniformly in the interval $[0,0.02]$ of admissible \emph{pump rates}. The neural network has two fully connected layers of 64 nodes each, activated by ReLU functions. The training was performed using Adam Optimizer with a learning rate of 0.001. We train the agent with 100 episodes \textit{in-silico} by using the model \eqref{eq:simple_dynamics}. Each episode comprises 100 steps with each time step equal to 1 minute, which is the sampling time imposed by the constraints of the bioreactor.
The synthetic OD measures, $x$, are generated by integrating \eqref{eq:simple_dynamics} with a smaller time step of $0.1 \, \mathrm{min}$, to simulate the continuous-time dynamics of the cells accurately. The results of the cumulative reward are shown in Figure \ref{fig:Pipeline_Deploy}.b.

Once a synthetically trained DQN had been obtained, we implemented the control strategy in real-time to regulate the OD inside a Chi.Bio. The bioreactor contained a culture of \textit{E. coli} strain hosting a plasmid with a genetic toggle switch.
The time evolution of the OD in Figure \ref{fig:Pipeline_Deploy}.c demonstrate the controller's effectiveness in successfully reaching and maintaining the desired set-point, set at 0.5. This was achieved with an average settling time of 10 minutes, indicating the controller's ability to efficiently and rapidly stabilize the system.\\
Next, we test the performance of the proposed controller to change the desired OD and its robustness to variation of the temperature of the culture affecting the intrinsic growth rate of the cells.


\subsection{In vivo performance and robustness assessment}
\begin{figure}
\centering

\includegraphics[width=1\linewidth]{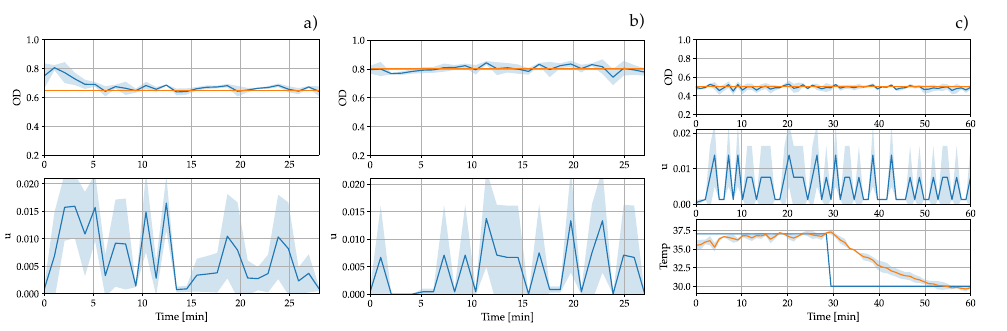}
\caption{Performance and robustness analysis: Panels a) and b) show the results of the OD regulation for reference of 0.65 and 0.8 respectively. The top subpanels depict the time evolution of the OD while the bottom subpanels depict the control input computed by the DQN-based controller. 
Panel c) shows the results of the OD regulation for reference 0.5 when subject to temperature change. After 30 minutes, the temperature is switched from 37°C to 30°C.
Solid lines represent the average evolution of state and input, while the shaded areas represent the standard deviation obtained over the three \textit{in vivo} experiments. The orange dashed line indicates the set point $\bar{x}$ of 0.5.
}
\label{fig:DQN_Performance_Robustness}
\end{figure}
We tested the performance of the \emph{sim-to-real} DQN controller conducting several experiments with the Chi.Bio. After a recovery phase, in which the cells were left to grow with abundant nutrients, we diluted the sample to a target OD value of 0.8. We kept this reference OD value for 30 minutes. After that, we switched to a target value of 0.65 and after 30 minutes we switched it to 0.5 for another 30 minutes. The experiments were replicated three times, for each desired value of OD of 0.8, 0.65, and 0.5. 
The average of the controlled OD together with the standard deviations for each experimental scenario are shown in Figure \ref{fig:DQN_Performance_Robustness}a-b.

Moreover, we assessed the robustness of the DQN controller to temperature variation, which directly affects the growth rate of the cells. Indeed, with respect to the nominal condition of 37°C, we experimentally observed a decrease of $10 \% $ of the growth rate when the temperature was changed to 30°C.
The test was conducted by starting the experiments at 37°C, regulating the OD to a target value of 0.5 for 30 min. After that, we switched the temperature of the Chi.Bio to 30°C.
The results of this experiment are reported in Figure \ref{fig:DQN_Performance_Robustness}.c. 
It can be noticed that the controller successfully maintains the OD to the desired value despite the perturbation in the intrinsic growth rate of the cell due to the change of the temperature capturing its robustness.

\section{Control benchmarks and comparison}
In what follows, we compare the proposed learning-based controller to other controller types often used in synthetic biology applications for the regulation of biochemical processes, namely the Proportional Integral (PI) controller and the Model Predictive Controller (MPC).
The PI controller considered here is already embedded in the Chi.Bio \citep{steel2019chi}, while the MPC has been specifically designed here for the sake of comparison.
Based on the error between the desired OD value and the measured OD, the PI controller evaluates a proportional action and two integral actions: one classical action to reject steady-state error, and another one to compensate for the effect of faulty gaskets in the pumps. 
%
The MPC evaluates what value of the control input must be applied to the system by solving an optimization problem at each control cycle.
Specifically, at each time step, it solves an optimization problem on a finite prediction horizon of length $T_h = 5\,\mathrm{min}$ searching for the policy that minimizes the cost function:
\begin{equation}
J = \sum_{k=0}^{N-1}{ c_k} + V_F(x_N) ,
\end{equation}
where the cost term $c_k$ is defined as:
\begin{equation}
c_k = \begin{cases}100 & \quad  \text{if }  u\notin [0,0.02]\\
(x_k-\bar{x})^2 & \quad  \text{otherwise }
\end{cases} ,
\end{equation}
so as to penalize also the violation of the constraints on the actuators; the final cost being defined as $V_F(x_N) = (x_N-\bar{x})^2$.
The algorithm employs model \eqref{eq:simple_dynamics} to run the optimization problem, which is solved by means of a particle swarm optimizer \citep{bonyadi2017particle}.
The control input obtained as a solution to this optimization problem is then applied to the real system in the next time interval of 1 min (i.e. the control horizon $T$).

\subsection{Comparison}

\begin{figure}
\centering
\includegraphics[width=1\linewidth]{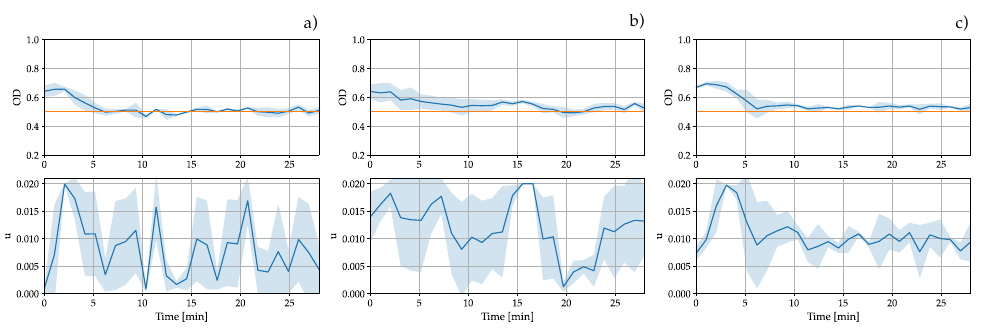}
\caption{Controllers ' comparison: The results of the OD regulation for reference of 0.5. The top panels depict the time evolution of the OD while the bottom panels depict the control input computed by a) DQN, b) PI, c) MPC, respectively.
Solid lines represent the average evolution of state and input, while the shaded areas represent the standard deviation obtained over the three realizations of the experiment. The orange dashed line indicates the set point $\bar{x}$ of 0.5. }
\label{fig:Comparison}
\end{figure}

To assess quantitatively the performance of the control algorithms we used two integral metrics, namely the Integral Squared Error (ISE) and the Integral Time Absolute Error (ITAE), that provide a quantitative measure of the transient and static performance, respectively. More precisely, the ISE and ITAE are defined as \citep{fiore2016vivo,guarino2020balancing}:
\begin{equation}
\mathrm{ISE} = \frac{1}{T} \int_0^t (\bar{x}-x(\tau))^2 d\tau, \quad \mathrm{ITAE} = \frac{1}{T} \int_0^t \tau|\bar{x}-x(\tau)| d\tau ,
\end{equation}
where $\bar{x}$ is the desired density and $T$ is the control horizon. 
The outcome of the experiments shown in Figure~\ref{fig:Comparison} confirms the ability of all the controllers to regulate the density of the population of interest. Furthermore, Table \ref{table:2} shows the controller's performances, comparing the learning-based control strategy with the PI and the MPC, confirming the viability of a \textit{sim-to-real} paradigm in a biological setting. Note that all the controllers have comparable performances with the DQN offering comparable performance and robustness to those of the MPC (see Table \ref{table:2}) despite the use of a sim-to-real approach based on an uncertain, simple model of the bacterial growth dynamics.
%
\begin{table}[t]
\begin{center}
\begin{tabular}{|c c c c|} 
 \hline
   & DQN & PI & MPC  \\
 \hline
\multicolumn{4}{ |l| }{\emph{Reference 0.8}} \\
 ISE   & $0.039$  & $0.046$& $0.035$\\
 ITAE  & $12.21$ & $12.38$ & $10.44$ \\ 
 \hline
 \multicolumn{4}{ |l| }{\emph{Reference 0.65}} \\
 ISE  & $0.032$ &$0.039$ & $0.092$\\
 ITAE & $11.90$ & $10.50$  & $9.38$\\  
 \hline
\multicolumn{4}{ |l| }{\emph{Reference 0.5}} \\
 ISE  & $0.051$ &$0.111$ & $0.117$ \\
 ITAE & $7.43$ & $11.49$ & $12.98$\\  
 \hline
\end{tabular}
\quad
\begin{tabular}{|c c c c|} 
 \hline
   & DQN & PI & MPC  \\
 \hline
\multicolumn{4}{ |l| }{\emph{Temperature 37°C}} \\
 ISE   & $0.045$  & $0.033$& $0.032$\\
 ITAE  & $12.42$ & $9.65$ & $11.60$ \\ 
 \hline
 \multicolumn{4}{ |l| }{\emph{Temperature 30°C}} \\
 ISE  & $0.031$ &$0.025$ & $0.042$\\
 ITAE & $11.84$ & $9.12$  & $11.54$\\  
 \hline
\end{tabular}
\caption{Control performance and robustness comparison via the metric ISE and ITAE}
\label{table:2}
\end{center}
\end{table}

\section{Discussion}

We regulate the growth of an \textit{E. coli} population in a small turbidostat by using a machine learning-based external control approach to regulate OD measures. To address the data efficiency issue that could render the algorithm impractical for synthetic biology applications, we adopted and experimentally validated the use of a \emph{sim-to-real} paradigm. In particular, the policy is initially acquired through training with a mathematical model of cell growth. This model was parametrized using a limited number of experiments, allowing us to address the substantial data requirement needed during the training phase. Subsequently, we validated it through \textit{in vivo} experimental testing.
Our experiments confirm the feasibility of this approach, demonstrating that it is possible to close the gap between simulations and experiments with a learning-based controller that can effectively regulate population density in a compact bioreactor like the Chi.Bio during \textit{in vivo} experiments. 
Starting from the results presented here, future work will be focused on the development of a learning-based controller leveraging the difference in growth rates of two different cell populations to regulate their relative densities inside a bioreactor, a much harder problem to solve with the more traditional approaches.

\acks{We express our sincere gratitude to the TIGEM Institute and Scuola Superiore Meridionale  for their support and resources, which contributed to the successful completion of this scientific paper. }
\newpage
\bibliography{references}

\end{document}